\documentclass[structabstract]{aa}  
\newcommand{\vv}{\mbox{\bf {v}}}

\newcommand{\vnh}{\hat{\mbox{\bf {n}}}}

\usepackage{multirow}
         
\def\plotancho#1{\includegraphics[width=18cm]{#1}}
\usepackage{graphicx}
\usepackage{txfonts}
\usepackage{natbib}
\usepackage{epsfig}

\begin{document}
   \title{Missing baryons, bulk flows and the  E-mode polarization of the Cosmic Microwave Background}


   \author{Carlos Hern\'andez--Monteagudo
          \inst{1}
          \and
          Rashid A. Sunyaev\inst{1,2}\fnmsep
                    }

   \institute{Max Planck Institut f\"ur Astrophysik, 
              Karl Schwarzschild Str.1, D-85741,
              Garching bei M\"unchen, Germany\\
              \email{chm@mpa-garching.mpg.de}
          \and
          	Space Research Institute, Russian Academy of Sciences, Profsoyuznaya 84/32, 117997 Moscow, Russia
             \\
            \email{sunyaev@mpa-garching.mpg.de}
                          }

   \date{Received ; accepted }

 
 \abstract
   { Most of the missing baryons are found in slightly overdense structures like filaments and superclusters, but to date most of them have remained hidden to observations. Similarly, the linear cosmological perturbation theory predicts the
 existence of extended bulk flows seeded by gravitational
 attraction of linear potential wells, but again these remain undetected.}
  {Unveil the signature of bulk flows of the missing baryons in upcoming Cosmic Microwave Background (CMB) data.}
   {If the peculiar
 motion of galaxy groups and clusters indeed resembles that of the surrounding baryons, then the
 kinetic Sunyaev-Zel'dovich (kSZ) pattern of those massive halos should be closely correlated to the 
kSZ pattern of all surrounding electrons. Likewise, it should also be correlated to the CMB E-mode polarization field generated via Thomson scattering after reionization. We explore the cross-correlation of the kSZ generated in groups and clusters to the all sky E-mode polarization in the context of upcoming CMB experiments like Planck, ACT, SPT or APEX. }
   {We find that this cross-correlation is effectively probing redshifts below $z=3-4$ (where most of baryons are missing), and that it arises in the very large scales ($l<10$). The significance with which this cross-correlation can be measured depends on the Poissonian uncertainty associated to the number of halos where the kSZ is measured and on the 
accuracy of the kSZ estimations themselves. Assuming that Planck can provide a cosmic variance limited E-mode polarization map at $l<20$ and  S/N $\sim 1$ kSZ estimates can be gathered for all clusters more massive than $10^{14} M_{\odot}$, then this cross-correlation should be measured at the  2--3 $\sigma$ level. Further, if an all-sky ACT or SPT type CMB experiment provides kSZ measurements for all halos above $10^{13} M_{\odot}$, then the cross-correlation signal to noise (S/N) ratio should be at the level of 4--5. A detection of this cross-correlation would provide direct and definite evidence of bulk flows and missing baryons simultaneously.}
   {}

 \authorrunning{Carlos Hern\'andez--Monteagudo \& Rashid A. Sunyaev}
 \titlerunning{Missing baryons, bulk flows and the E-mode CMB Polarization}
 
 \maketitle
%

\begin{figure*}
\centering
\plotancho{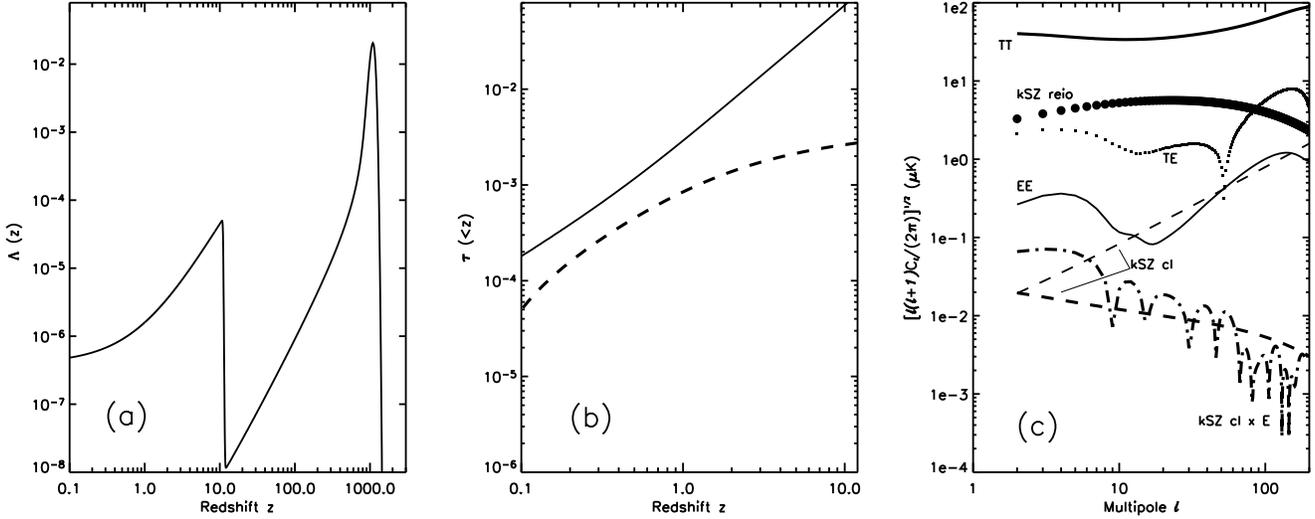}
\caption[fig:fig1]{ ({\it Left panel, a}): Visibility function versus redshift
under WMAP5 cosmogony.  ({\it Middle panel, b}): Thomson optical depth versus
redshift due to all electrons (solid line) and to those electrons contained in
halos more massive than $5\times 10^{12} M_{\odot}$ (dashed line). 
({\it Right panel, c, square root of angular
power spectra shown, note the $\mu$K units}). [Thick solid line:] Total CMB
temperature anisotropy amplitude. [Thin solid line:] Thomson scattering
induced fluctuations during and after reionization. [Thick dashed line:]
Correlation term of the Thomson scattering induced fluctuations generated in
galaxy clusters. [Thin dashed line:] Poisson term of the Thomson scattering
induced fluctuations in galaxy clusters. [Dot-dashed line:] Cross-correlation
between the total E-mode polarization  (thin solid line) and the correlated part of
the kSZ generated in galaxy groups and clusters (thick dashed line).}
\label{fig:fig1}
\end{figure*}

\section{Introduction}

One of the fundamental predictions of the standard cosmological model is that baryons (and matter in general) should be moving in extended bulk flows generated by the gravitational pull of large scale overdense regions. Despite they are potentially useful in terms of cosmological studies \citep{atrioksz1,atrioksz2,atrioksz3}, none of those bulk flows have been detected to date \citep{sarah}, partially due to the difficulty of detecting baryonic matter in the local universe. Indeed, the visible baryonic matter in the various frequency ranges amounts only to one ninth of the total budget of baryons predicted by the cosmological model \citep[][this is the so called {\it missing baryon} problem]{fukpeebles}. 
All these baryons must be found in the present Universe, since they have been detected at large ($z\simeq 1,100$) and intermediate redshift ($z\sim 6$), by means of Cosmic Microwave Background (CMB) \citep{wmap5} and Lyman-$\alpha$ forest \citep{lyman1,lyman2} observations. However, no conclusive observational evidence has  been found at low redshift.  Some indications have been derived from the observation of absorption lines in the direction of quasars (see \citet{qspec1,qspec2} and references therein), or a soft X-ray excess around clusters, \citep[e.g.,][]{xr1,xr2}. \citet{cenost1,cenost2} concluded that most of the baryons are in a warm phase (WHIM), defined by the temperature range $T \in [10^5, 10^7]$K. Whether most of this gas is in a diffuse phase or in small, unresolved, collapsed objects is still an open question. Recently, \citet{rgs1, rgs2} detected a non-Gaussian cold spot when looking at the CMB radiation in the direction of Corona Borealis, and discussed its interpretation in the context of the thermal and the kinetic Sunyaev-Zel'dovich effects. The thermal Sunyaev-Zel'dovich effect \citep[tSZ,][]{tSZ} expresses the distortion that the black body CMB spectrum undergoes due to Compton scattering on hot electrons. On the other hand, the kinetic Sunyaev-Zel'dovich \citep[kSZ,][]{kSZ} describes the Thomson scattering between CMB photons and electrons in which there is no energy exchange (and therefore it introduces thermal or frequency independent brightness temperature fluctuations). In the case of Corona Borealis, it was suggested that a face on filament could give rise, via either tSZ or kSZ, to the pattern seen in the CMB maps. However, it was shown in \citet{letterdtsz} that the tSZ is not a useful tool to track the baryons outside the largest collapsed halos: about $\sim 80$ \% of the tSZ luminosity is generated in the most massive collapsed structures, and the $\sim$ 50\% of baryons located in underdense or slightly overdense regions give rise to only $\sim$ 5\% of the total tSZ luminosity. The kSZ is most promising, since it does not require large gas pressure, but relative velocities with respect to the CMB instead (which is itself a prediction of the model). However, the drawback is that the kSZ temperature fluctuations do not depend on the frequency, and therefore they cannot be easily distinguished from the intrinsic CMB temperature fluctuations. 

In this work we profit the anisotropic nature of the Thomson scattering (for which linear polarization is introduced in the CMB), and use it as a tool to unveil the bulk flows of the missing baryons in the low redshift universe, ($z < 2-3$). For that, we need to assume that the peculiar motion of clusters and groups are correlated to the velocities of the surrounding matter (just as the linear theory predicts),
and use measurements on halos as a kSZ template for all baryons. In Section (\ref{sec:reio}) we compute the kSZ and the polarization generated during and after reionization, and the cross-correlation between both. In Section (\ref{sec:halos}) we compute the kSZ produced by the halo population and the signal to noise (S/N) ratio for its cross-correlation to the all-sky E-mode polarization. We discuss our results and conclude in Section (\ref{sec:conclusions}). Throughout this paper, we shall be using a cosmological parameter set corresponding to the WMAP5 + $\Lambda$CDM model, \citep{wmap5}.

\section{The large scale secondary anisotropies after reionization}
\label{sec:reio}

According to WMAP data, the first stars must have reionized the IGM at around
$z_{reio}\simeq 11$, \citep{wmap5}. Since then, CMB photons and electrons have
interacted again via Thomson scattering.  As during recombination, this
scattering preserves the black body spectrum of the CMB, but however changes
the direction of the CMB photons, and, in the presence of a non-vanishing
quadrupole of the CMB intensity, in also induces linear polarization. This
process is the cause of a partial blurring of the anisotropies generated at
the surface of last scattering and also the generation of new anisotropies at
scales comparable to the horizon at $z=z_{reio}$ ($\theta \sim 20\degr$ or
$l\sim 10$). (One should also consider that, if reionization proceeded first
in those most overdense regions hosting the first stars, then it must also
have left a patchy signature on the CMB small scale anisotropies. However, in
this work we shall neglect these small angle fluctuations).  The temperature anisotropies
introduced in the CMB intensity were caused by the peculiar motions of the
scatterers (electrons) with respect to the CMB, (i.e., the kSZ effect). A simple expression
for these temperature anisotropies is given by the following integral of the
projection of the peculiar velocity along the line of sight $\vnh$:
\begin{equation}
\frac{\delta T}{T_0} (\vnh) \simeq \int_{\eta_{reio}}^{\eta_0} d\eta \; 
\Lambda(\eta ) 
 \left( -\frac{\vnh\cdot \vv_e (\vnh, \eta)}{c}\right),
\label{eq:dtt_lin1}
\end{equation}
(note that we are neglecting the few percent contribution of the polarization
terms).  As in \citet{selzal96}, in this equation $\eta$ is the conformal time
(related to the coordinate time via $d\eta = dt / a(t)$, with $a$ the
cosmological scale factor), and $\eta_{reio}, \eta_0$ are its values at
reionization and present, respectively. Dots will denote derivatives with
respect to the conformal time.  The (proper) electron peculiar velocity is
given by $\vv_e (\eta)$, and its Fourier counterpart can be related to the
underlying linear density perturbation field by means of the continuity
equation. These peculiar velocities describe the large scale {\em bulk flows}
of matter that are generated by the linear perturbations in the matter density
field. The visibility function is defined as $\Lambda (\eta ) = \dot{\tau}
(\eta ) \exp{\left( -\tau(\eta)\right)}$, with $\tau (\eta ) $ the Thomson
optical depth given by $\tau (\eta ) \equiv \int_{\eta}^{\eta_0} d\eta'
\;a(\eta') \sigma_T n_e(\eta')$. In this expression, $n_e(\eta)$ the
background electron number density and $\sigma_T$ the Thomson cross section.
The Thomson opacity is given by $\dot{\tau} (\eta ) \equiv a(\eta) \sigma_T
n_e(\eta )$.  The visibility function $\Lambda(\eta )$ denotes the probability
that a CMB photon was last scattered at the epoch given by $\eta$. In panel
(a) of Figure (\ref{fig:fig1}) it can be seen that the visibility function
peaks at recombination redshift $z\simeq 1,100$ (as first predicted by
\citet{sunyaev70}), although it has a second peak at $z = z_{reio}\simeq 10$,
denoting the new Thomson scattering taking place after the stars reionize the
IGM. Note that the optical depth behaves differently, increasing steadily upto
reionization redshift (solid line of panel (b) in Figure (\ref{fig:fig1})).

The anisotropic nature of the Thomson scattering introduces linear
polarization as long as the CMB intensity has a non-vanishing local
quadrupole. In CMB studies, it is customary to map the standard $Q$ and $U$
polarization Stokes parameters into the (rotation invariant) $E$ and $B$ modes
\citep{zaldaseljakpol,kamionpol}, which behave with opposite parity properties
versus spatial reflections. Pure Thomson scattering should produce no $B$ mode
polarization, and hereafter this mode will be neglected. Since the
evolution of the intensity and E-mode polarization anisotropies are coupled
by the sum of the intensity quadrupole and the polarization monopole and quadrupole,
 both quantities will be correlated, and actually
this cross-correlation has already been measured by WMAP, \citep{wmap5}.

Using a modified version of a standard Boltzmann code (like, e.g., CMBFAST
\citep{selzal96}) it is straightforward to compute the temperature and E-mode
polarization anisotropies generated during and after reionization, and the cross-correlation
between them.The results,
computed under WMAP5 cosmology \citep{wmap5}, are displayed in Figure
(\ref{fig:fig1}c). The total intensity (TT) and E polarization (EE) angular
power spectra are displayed by the thick and thin solid lines, respectively,
whereas the cross-correlation between the intensity and E polarization is
given by the dotted line (TE). The temperature anisotropies induced by the kSZ
during and after reionization is given by the filled circles, and the
generation of polarization at the same cosmic epochs is the responsible for
the low $l$ ($l\sim 5-8$) bump in the EE spectrum. Although the
WMAP\footnote{WMAP URL site: {\tt http://map.gsfc.nasa.gov}} mission has
barely measured it, the upcoming Planck\footnote{Planck's URL site: \\
{\tt
http://www.rssd.esa.int/index.php?project=planck}} satellite should be able to provide an accurate estimate of the EE
angular power spectrum (and this shall be our assumption hereafter).

\section{The clusters as probes of bulk flows}
\label{sec:halos}

As mentioned in the Introduction, there are two major aspects of our
understanding of the local Universe that still are pending for confirmation:
one is the detection of the eight ninths of baryons that have not yet been
found \citep{fukpeebles}, the another is the measurement of the cosmological
bulk flows. With the advent of high resolution and high sensitivity CMB
experiments like ACT \citep{ACT}, SPT \citep{SPT} or APEX \citep{APEX}, bulk flows have prospects
to be detected by looking at the kSZ in galaxy clusters. These structures
contain large reservoirs of gas, which constitute sources for relatively large
values of optical depth, ($\tau \sim 10^{-4},10^{-3}$). Nevertheless, one must
have in mind that the kSZ effect is not the only physical mechanism to be
found in clusters. Indeed, given the fact that, in these halos, gas can reach
temperatures at the level of a few KeV, inverse Compton scattering in which
hot electrons transfer energy to CMB photons (tSZ)
 becomes of relevance as well. However, the brightness temperature
fluctuations introduced by this effect cross from negative to positive values at a
frequency close to 218 GHz. For this reason, this frequency becomes an ideal
window for searching for the kSZ in galaxy clusters.

\begin{figure*}
\centering
\plotancho{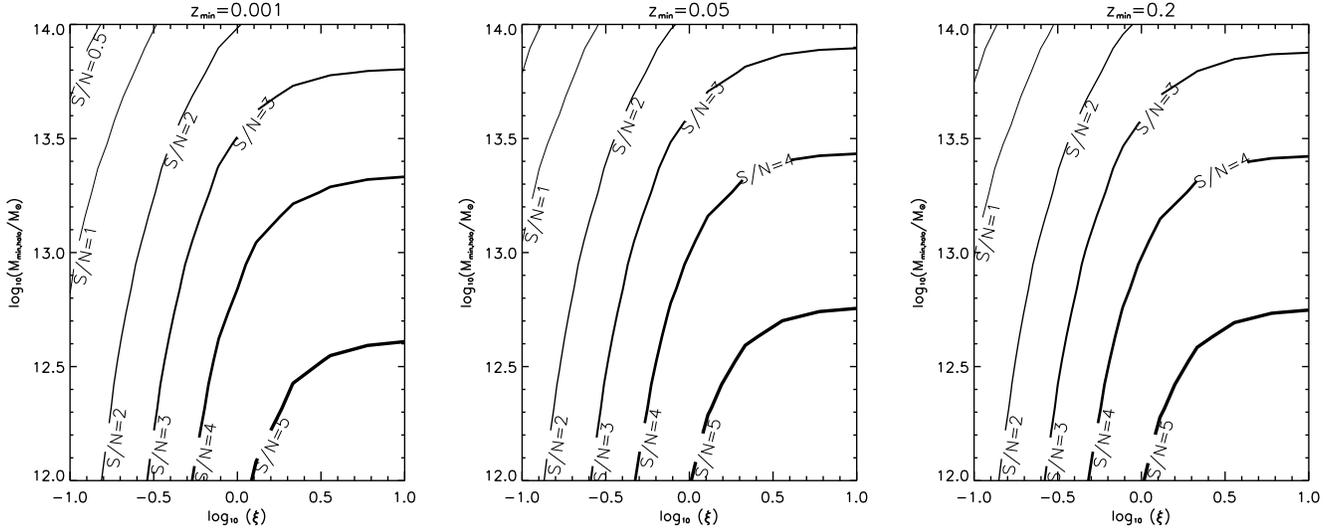}
\caption[fig:fig2]{ The three panels display the total S/N ratio for the correlation of the E-mode polarization with the low redshift ($z<3-4$) kSZ generated by bulk flows in galaxy groups and clusters. Horizontal axes display the decimal logarithm of the individual group/cluster kSZ measurement S/N ratio ($\xi$), whereas vertical axes display the decimal logarithm of the minimum mass of groups/clusters considered. Due to Poisson/shot noise, a low choice for $z_{min}$ makes the high total S/N regions shrink, and this also happens if $z_{min}$ is chosen too big and too much information is unused. The optimal minimum redshift is around $z_{min} \simeq 0.05$, and the maximum S/N ratio achievable for perfect E and kSZ surveys is $\sim 5.8$. Note that we are assuming $f_{sky} = 1$.
}
\label{fig:fig2}
\end{figure*}

The kSZ effect generated in the galaxy cluster population has been studied
extensively \citep[][etc]{peel06,kszchm,arthur2,arthur1}. In particular, in
\citet{kszchm} an analysis of the correlation properties of the cluster
peculiar motion was provided. They found that the typical correlation length
of these peculiar velocities (i.e., the typical size of a bulk flow) is
typically 20 $h^{-1}$Mpc in comoving units. This means that a great fraction
of the gas that is surrounding the clusters is actually {\em comoving} with
it. This can be envisioned as flows of matter (containing clusters, groups and
diffuse gas) falling into the (future) supercluster's potential well, and
generating a kSZ dipole pattern at its center  as
different flows merge in that region, \citep{diaferiosunyaev}. This is related to the basic assumption
of this work: {\em peculiar motion of baryons in a given region is well
described by the peculiar motion of the largest galaxy groups and clusters
within the same region}. When computing the all sky kSZ angular power spectrum
generated by clusters, \citet{kszchm} found that although the this could be
decomposed in four terms, only two of them were giving rise to most of
the power: a correlation term (generated by the bulk flows) was dominant at
the very large angular scales, whereas the anisotropy induced by the Poisson
statistics would dominate at smaller scales. Both correlation and Poisson
terms are shown by the thick and thin dashed lines in Figure
(\ref{fig:fig1}c). The former has been computed using a modified version of
the CMBFAST code where the Thomson opacity generated in clusters at a given
$\eta$ reads \citep{kszchm}:

\begin{equation}
\dot{\tau} = a(\eta)\sigma_T\;\int_{M_{min}}^{M_{max}} dM\; \frac{dn}{dM} N_e(M)\; b_v.
\label{eq:opac_ksz_cl}
\end{equation}
The symbol $N_e(M)=M/(m_p f_Y)\Omega_b/\Omega_m$ denotes the number of
electrons in halo of mass $M$, with $m_p$ the proton mass, $f_Y=1-Y/2$ the
Helium fraction correction factor and $\Omega_b$, $\Omega_m$ the baryon and
matter density parameters, respectively. The halo mass function $dn/dM$ is
approximated by that of \citet{ST}.  Motivated by \citet{peel06},
the velocity bias was approximated by $b_v = 1.3$ for all halos (although the
particular choice of this parameter has very little impact on our results, as
we shall see below). We set $M_{max}=10^{16}$ M$_{\odot}$, and compared our
results for different choices of the minimum mass $M_{min}$. In the middle panel of 
Figure (\ref{fig:fig1}) the dashed line shows the optical depth generated by halos more
massive than $5\times 10^12$ $M_{\odot}$: most of the contribution ($\sim 60$\%) is coming from
$z<3$, and the contribution from (the more massive) galaxy groups and clusters is shifted even
 further into the low $z$ range.
On the other hand, the Poisson term can easily be found to be
\[
C_l^{kSZ,\xi} = \int_{\eta_{reio}}^{\eta(z_{min})}d\eta\; \eta^2\;\int_{M_{min}}^{M_{max}}dM\; \frac{dn}{dM} 
  \left(\frac{\sigma_T N_e(M)b_v \sigma_v(M,\eta)}{d_{Ang}^2(\eta )}\right)^2\; 
  \]
\begin{equation}  
 \phantom{xxxxxxxxxxxxxxxxxxxxxxxxxxxx} \times \;
  \left(1+\frac{1}{\xi^2}\right),
\label{eq:cl_poisson}
\end{equation}
where $d_{Ang} (\eta)$ denotes angular diameter distance and
$\sigma_v(M,\eta)$ is the time dependent rms line of sight peculiar
velocity of an object of mass $M$,
\begin{equation}
\sigma_v^2 (M, \eta) = \frac{1}{3}\left( H(\eta) \biggl| \frac{d{\cal D}_{\delta}}{dz}\biggr|\right)^2
   \int dk\; k^2\; \frac{P_m(k)}{2\pi^2 k^2} \bigl| W(kR(M))\bigr|^2.
\label{eq:sigma_v}
\end{equation}
The present linear matter power spectrum is given by $P_m(k)$, the linear density
growth factor by ${\cal D}_{\delta}$, the Hubble parameter by $H(\eta)$, and
the Fourier window function of a top hat filter corresponding to an object of
mass $M$ by $W(kR(M))$, (see \citet{kszchm} for details).  Note that in
equation (\ref{eq:cl_poisson}), $\xi$ denotes the signal to noise ratio of the
kSZ detection in a single cluster/group. As noise we regard {\em any} signal
present at the cluster position that is not due to the correlated component of
the kSZ (e.g., kSZ caused by thermal velocities, tSZ residuals, intrinsic CMB,
radio source emission, IR source emission, instrumental noise, etc). Note as
well the low redshift limit given by $z_{min}$ in equation
(\ref{eq:cl_poisson}). Therefore, this equation accounts for both the
Poissonian statistics {\em and} the error in the kSZ recovery at cluster
positions. Only in the case $\xi \rightarrow \infty$ we recover the pure
Poissonian term.

We are interested in the correlated component of the cluster kSZ, i.e., in the
velocity component that should be comoving with the surrounding matter. If
this is indeed the case, then the pattern of the kSZ in clusters must be very
similar to that of the kSZ generated by {\em all} electrons (filled circles in
Figure (\ref{fig:fig1})), and therefore it must be closely correlated to the
$E$ mode of the polarization anisotropies generated during and after
reionization (i.e., the low $l$ bump). The actual amplitude of this
cross-correlation will depend on the redshift range that a given set of halos
is probing, and the signal to noise (S/N) ratio of this cross-correlation for
a particular multipole $l$ is provided by
\begin{equation}
\left( \frac{S}{N} \right)_l^2 = \frac{(2l+1)\;f_{sky} \; \left( C_l^{kSZ,cl - E}\right)^2}{C_l^{EE}\left(C_l^{kSZ,cl} + C_l^{kSZ, \xi}\right) + \left( C_l^{kSZ,cl - E}\right)^2}.
\label{eq:s2ncls}
\end{equation}
The symbol $f_{sky}$ refers to the fraction of the sky covered by an
experiment, and $C_l^{kSZ,cl} $ and $C_l^{kSZ, \xi}$ denote, respectively, the
angular power spectrum of the correlated part and the noise of the kSZ
generated in galaxy clusters. The cross-correlation between the kSZ in halos and the E-mode is
given by $C_l^{kSZ,cl - E}$. If $f_{sky}=1$, the total S/N is given by the
sum of the squares of these quantities, i.e., $S/N = \left[\sum_l
(S/N)_l^2\right]^{1/2}$. It turns out that the actual amplitude of this S/N ratio is
strongly dependent on the relative amplitude of the Poisson term with respect
to the correlation term (cancelling any dependence on parameters like the
velocity bias $b_v$). Due to the $\eta^2/d_{Ang}^4\propto 1/\eta^2$
dependence in equation (\ref{eq:cl_poisson}), the Poisson term is less
important for higher values of the minimum redshift considered
$z_{min}$. Further, it depends linearly with the number density of objects, as
opposed to the correlation term, which depends on the number density
squared. Therefore, for smaller values of the threshold mass $M_{min}$ the
relative weight of the Poisson term versus the correlation one should
diminish. This is shown in Figure (\ref{fig:fig2}). It is worth to remark as well that
since this correlation arises around halos at low and moderate redshifts, the matter
behind this signal is {\em different} from that generating the {\em anti} cross-correlation
between CMB anisotropies and HI 21cm observations.

\section{Discussion and Conclusions}
\label{sec:conclusions}

The horizontal axis of Figure (\ref{fig:fig2}) corresponds to the S/N ratio to
which the kSZ can be determined in a single halo, i.e., $\xi$ in equation
(\ref{eq:cl_poisson}). The number of potential contaminants when measuring the
kSZ in a halo is relatively high: the tSZ has a definite spectral dependence,
and should be close to zero at $\nu = 218$ GHz, but however asymmetries in the
frequency response of the experiment together with relativistic effects
\citep{reltSZ} will inevitably leave some residuals. However, as shown in
\citet{kszchm}, its relative weight to the kSZ decreases for the most numerous
low mass halos, for which the relativistic tSZ corrections are also less
important. Internal motions within the halos (which, as found by
\citet{diaferio05} are only relevant for slow halos)
should average out when integrating over the total halo solid angle. The
uncorrelated component of the kSZ could be significant in those systems where
the velocity of halos are thermalized; however this almost never happens for
the galaxy groups and clusters being targeted here, since superclusters are structures
that are not yet relaxed. Nevertheless, \citet{peel06} found that, in his
cosmological simulations, the peculiar velocities of haloes showed correlation properties
that differed substantially from linear theory predictions, even at high redshift. If this is indeed the case, then the issue would be whether the baryons surrounding halos would have their peculiar velocities affected in the same way or not. We shall address this issue in future work.
The impact of IR and radio sources is {\it a priori}
significant, provided that these sources are correlated to the halos where the
kSZ is produced. However, multifrequency observations should provide a good
handle on their spectral indexes, and therefore their
contribution at the 218 GHz channel should be characterized down to an
accuracy level to be realistically estimated only when real data becomes available.
The intrinsic CMB generated at the last scattering surface is
another potential source of contamination. As shown in Figure (6) of
\citet{kszchm}, its residuals remain in the few $\mu$K level if the
cluster/group size is not larger than 2-3 arcmins, which should be the case
for most of clusters and groups in the sky. Therefore this source of
contamination is likely to be less relevant than the rest.

The importance of each of these contaminants will determine which range of the $\xi$ horizontal
 axis in Figure (\ref{fig:fig2}) one should look at. Due to the largest amplitude of Poissonian fluctuations at low values of $z_{min}$, it is convenient not to consider too nearby clusters and groups. From this figure it is possible to find at which values of $\xi$ the S/N ratio starts being dominated by Cosmic Variance (horizontal {\it plateau} of the contour levels at high $\xi$). Likewise, the steep vertical descent of these contours at low $\xi$ reflects the limit where S/N is exclusively limited by the accuracy of the kSZ measurements.The requirements for a convincing detection of the signature of the local baryon bulk flows (say $S/N > 3$)
 involve either a large value of $\xi$ or a low value of $M_{min}$, but nevertheless these are not unrealistic. For instance, the Planck satellite should provide a cleaned CMB map in a very large fraction of the sky, and with it, estimations of the kSZ at the cluster and group positions. These estimates should mostly be limited by the beam size and the un-substracted point source emission.
By looking at the Compton distortion $y$ map (obtained from multifrequency observations of the tSZ),
one should be able to find the position of not only the largest clusters, but also of those objects that lie closer to the confusion limit, and whose signal can be picked up statistically via the cross-correlation. If this $y$-map provides the position and the cleaned CMB map the kSZ estimates (of at least  $\xi \sim 1$) for all clusters more massive than $10^{14} M_{\odot}$, then Planck alone would already yield a total S/N between 2 and 3.  Likewise, upcoming Dark Energy surveys like Pan-STARRS\footnote{Pan-STARRS' URL site: {\tt http://pan-starrs.ifa.hawaii.edu/public/}}, DES \footnote{DES's URL site: {\tt http://www.darkenergysurvey.org/}} or PAU-BAO\footnote{PAU-BAO's URL site: {\tt http://www.ice.csic.es/research/PAU/}}, or X-ray surveys like eROSITA\footnote{eROSITA's URL site: {\tt http://www.mpe.mpg.de/projects.html\#erosita}}  should probe the relevant redshift range, and provide independently catalogs of group and cluster candidates. At the same time, high resolution CMB experiments like ACT, SPT or APEX would contribute with more accurate kSZ estimates. Furthermore, future B-mode polarization experiments should cover large fractions of the sky with very high sensitivity, and provide better understanding of the E-mode of the CMB polarization. One must remark that the maximum achievable value of the total S/N is relatively small (around 5.8), and therefore {\em a large sky coverage is required}, since most of the signal is arising at $l<10$. Along these lines, we must stress again that a {\em cosmic variance limited} EE power spectrum has been assumed, at least in those same large angular scales. Very soon, the Planck mission should be able to tell whether this is a sensible assumption or not.

\begin{acknowledgements} 
C.H.M. acknowledges useful conversations with D.Spergel.
\end{acknowledgements}

\bibliographystyle{aa}
\bibliography{Lit}

\end{document}